\documentclass[letterpaper,twocolumn,twoside]{IEEEtran} 

\usepackage[utf8]{inputenc} 
\usepackage[T1]{fontenc}    
\usepackage{cite,hyperref}       
\usepackage{url}            
\usepackage{booktabs}       
\usepackage{amsfonts}       
\usepackage{nicefrac}       
\usepackage{microtype}      

\usepackage{graphicx}
\usepackage{psfrag}
\usepackage{amsmath, amssymb, amsthm}
\usepackage{algorithm,algorithmic}
\usepackage{bbm}
\usepackage{etoolbox}
\usepackage{enumerate}
\usepackage{tikz}
\usetikzlibrary{shapes,arrows}
\usetikzlibrary{patterns}
\usepackage{pgfplots}
\usepackage{xspace}
\usepackage{enumerate}
\usepackage{mathtools}
\usepackage{epstopdf}

\newtoggle{conference}
\togglefalse{conference} 

\newtheorem{theorem}{Theorem}

\usepackage[normalem]{ulem} 

\def\beq{\begin{equation}}
\def\eeq{\end{equation}}
\def\beqa{\begin{eqnarray}}
\def\eeqa{\end{eqnarray}}
\def\beqan{\begin{eqnarray*}}
\def\eeqan{\end{eqnarray*}}

\def\R{{\mathbb{R}}}

\def\argmin{\mathop{\mathrm{arg\,min}}}
\def\argmax{\mathop{\mathrm{arg\,max}}}

\def\Diag{\mathop{\mathrm{Diag}}}

\DeclareMathOperator{\Tr}{tr}
\DeclareMathOperator{\var}{var}

\DeclareMathOperator{\Cov}{Cov}
\def\x{\times}

\def\bhat{\widehat{b}}

\def\qhat{\widehat{q}}

\def\xhat{\widehat{x}}

\def\Exp{\mathbb{E}}

\def\Tm1{T\! - \! 1}
\def\tm1{t\! - \! 1}
\def\tp1{t\! + \! 1}
\def\km1{k\! - \! 1}
\def\kp1{k\! + \! 1}

\newcommand{\zero}{\mathbf{0}}

\newcommand{\dbf}{\mathbf{d}}

\newcommand{\qbf}{\mathbf{q}}

\newcommand{\rbf}{\mathbf{r}}

\newcommand{\ubf}{\mathbf{u}}

\newcommand{\wbf}{\mathbf{w}}

\newcommand{\xbf}{\mathbf{x}}
\newcommand{\xbfhat}{\widehat{\mathbf{x}}}

\newcommand{\ybf}{\mathbf{y}}
\newcommand{\zbf}{\mathbf{z}}

\newcommand{\Abf}{\mathbf{A}}
\newcommand{\Bbf}{\mathbf{B}}

\newcommand{\Cbf}{\mathbf{C}}
\newcommand{\Dbf}{\mathbf{D}}

\newcommand{\Ibf}{\mathbf{I}}

\newcommand{\Qbf}{\mathbf{Q}}

\newcommand{\Sbf}{\mathbf{S}}
\newcommand{\Ubf}{\mathbf{U}}
\newcommand{\Vbf}{\mathbf{V}}

\def\betabf{{\boldsymbol \beta}}

\newcommand{\thetabf}{{\boldsymbol{\theta}}}
\newcommand{\thetahat}{{\widehat{\theta}}}
\newcommand{\thetabfhat}{{\widehat{\boldsymbol{\theta}}}}
\newcommand{\phibf}{{\boldsymbol{\phi}}}

\newcommand{\tran}{^{\text{\sf T}}}

\newcommand*\dif{\mathop{}\!\mathrm{d}} 

\title{Learning and Free Energies for Vector Approximate Message Passing}

\iftoggle{conference}
{
\twoauthors{Alyson K. Fletcher%
       \thanks{A.~K. Fletcher was supported in part by the NSF grant CCF-1254204 
           and in part by Office of Naval Research Grant N00014-15-1-2677\@.
           P. Schniter was supported in part by the NSF grant CCF-1527162.}}
  {Depts. of Stats., Math., and Elec. Eng.\\
   Univ. of California, Los Angeles CA 90095}
{Philip Schniter}
  {Dept. of Elec. \& Comp. Eng.\\
   The Ohio State Univ., Columbus OH 43210}
}
{
\author{
Alyson K. Fletcher and
Philip Schniter
\thanks{A.~K. Fletcher (email: akfletcher@ucla.edu) is with
    the Departments of Statistics, Mathematics, and Electrical Engineering,
    University of California, Los Angeles, CA, 90095.
    The work of A.~K. Fletcher was supported by the NSF under grant CCF-1254204.}%
\thanks{P.~Schniter (email: schniter.1@osu.edu) is with
    the Department of Electrical and Computer Engineering,
    The Ohio State University, Columbus, OH, 43210.
    The work of P. Schniter was supported by the NSF under grant CCF-1527162.}
}
}

\begin{document}

\maketitle

\begin{abstract}
Vector approximate message passing (VAMP) is a computationally simple approach to the recovery of a signal $\mathbf{x}$ from noisy linear measurements $\mathbf{y}=\mathbf{Ax}+\mathbf{w}$.
Like the AMP proposed by Donoho, Maleki, and Montanari in 2009, VAMP is
characterized by a rigorous state evolution (SE) that holds under certain large random matrices and that matches the replica prediction of optimality.
But while AMP's SE holds only for large i.i.d.\ sub-Gaussian $\mathbf{A}$, VAMP's SE holds under the much larger class: right-rotationally invariant $\mathbf{A}$.
To run VAMP, however, one must specify the statistical parameters of the signal and noise.
This work combines VAMP
with Expectation-Maximization to yield an algorithm, EM-VAMP, that
can jointly recover $\mathbf{x}$ while learning those statistical parameters.  The fixed points of the proposed EM-VAMP
algorithm are shown to be stationary points of a certain constrained free-energy, providing a variational interpretation of the algorithm.
Numerical simulations show that EM-VAMP is robust to highly ill-conditioned $\mathbf{A}$ with performance nearly matching oracle-parameter VAMP\@.
%
\end{abstract}


\section{Introduction} \label{sec:intro}

Consider the problem of estimating a random vector $\xbf$
from linear measurements $\ybf$ of the form
\beq \label{eq:yAx}
    \ybf = \Abf \xbf + \wbf, \quad
    \wbf \sim {\mathcal N}(\zero,\theta_2^{-1} \Ibf),
    \quad
    \xbf \sim p(\xbf|\thetabf_1),
\eeq
where $\Abf\in\R^{M\times N}$ is a known matrix,
$p(\xbf|\thetabf_1)$ is a density on $\xbf$ with
parameters $\thetabf_1$,
$\wbf$ is additive white Gaussian noise (AWGN) independent of $\xbf$, and $\theta_2 > 0$ is the noise precision
(inverse variance).
The goal is to estimate $\xbf$ along while simultaneously learning the unknown parameters
$\thetabf := (\thetabf_1,\theta_2)$ from the data $\ybf$ and $\Abf$.
This problem arises in Bayesian forms
of linear inverse problems in signal processing,
as well as in linear regression in statistics.

Even when the parameters $\thetabf$ are known,
exact estimation or inference of the vector $\xbf$ is intractable
for general priors $p(\xbf|\thetabf_1)$.
The \emph{approximate message passing} (AMP) algorithm \cite{DonohoMM:09} and its generalization \cite{Rangan:11-ISIT}
are powerful, relatively recent, algorithms that
iteratively attempt to recover $\xbf$.
These methods are computationally fast
and have been successfully applied to a wide range of problems, e.g.,
\cite{FletcherRVB:11,
Schniter:11,
SomS:12,
ZinielS:13b,
Vila:TSP:14,
KamilovGR:12,
Schniter:TSP:15,
Ziniel:TSP:15,
fletcher2014scalable
}.
Most importantly, for large, i.i.d., sub-Gaussian random matrices $\Abf$,
their performance can be exactly predicted by a scalar \emph{state evolution}
(SE) \cite{BayatiM:11,javanmard2013state} that provides testable conditions
for optimality, even for non-convex priors.
When the parameters $\thetabf$ in the model are unknown,
AMP can be combined with expectation maximization (EM) methods
\cite{krzakala2012statistical,vila2013expectation,KamRanFU:12-IT}
for joint estimation and learning.

As it turns out, the AMP methods \cite{DonohoMM:09,Rangan:11-ISIT} are
fragile with regard to the choice of the matrix $\Abf$, and can perform poorly outside the special case of zero-mean, i.i.d., sub-Gaussian $\Abf$.
For example, AMP diverges with even mildly non-zero-mean and/or mildly ill-conditioned $\Abf$ \cite{Vila:ICASSP:15}.
Several techniques have been proposed to improve the robustness of AMP
including damping \cite{RanSchFle:14-ISIT,Vila:ICASSP:15}, mean-removal \cite{Vila:ICASSP:15}, and sequential updating \cite{manoel2015swamp}, but these remedies have limited effect.

Recently, the Vector AMP (VAMP) algorithm \cite{rangan2016vamp}  was established as an alternative to AMP that is much more robust to the choice of matrix $\Abf$.
In particular, VAMP has a rigorous SE that holds under large right-rotationally invariant $\Abf$, i.e., $\Abf$ whose right singular-vector matrix is uniformly distributed on the group of orthogonal matrices.
VAMP can be derived in several ways, such as through expectation propagation (EP) \cite{Minka:01} approximations of belief propagation \cite{rangan2016vamp} or through expectation consistent (EC) approximation \cite{opper2004expectation,OppWin:05,fletcher2016expectation}. 
But the existence of a rigorous state evolution establishes it firmly in the class of AMP algorithms.


However, a shortcoming of the VAMP method \cite{rangan2016vamp} is that it 
requires that the parameters $\thetabf$ in the model \eqref{eq:yAx} are known.
In this paper, we extend the VAMP method to enable learning
of the parameters $\thetabf$ via Expectation-Maximization (EM) 
\iftoggle{conference}{\cite{NealHinton:98}}{\cite{DempLR:77,NealHinton:98}}.
We call the proposed method EM-VAMP\@.
As described below, exact implementation of EM requires estimating
the posterior density $p(\xbf|\ybf,\thetabfhat)$ for each parameter estimate
$\thetabfhat$.  This is computationally not possible for the model~\eqref{eq:yAx}.
EM-VAMP is instead derived using a technique from Heskes
\cite{heskes2004approximate} for combining EM with approximate inference
of the posterior.  Specifically, it is well-known that EM can be interpreted
as a method to minimize a certain energy function.  Here, we construct
an approximation of the EM cost function that we call the EM-VAMP energy function
and derive an algorithm to minimize this function.

Our main theoretical result shows that
the fixed points of the EM-VAMP method are local minima of the
EM-VAMP energy function and thus provide estimates of the
parameters $\thetabf$ and posterior density with a precise variational interpretation.
By including the parameter learning,
this result generalizes the fixed-point energy-function interpretation of
EC given in \cite{opper2000gaussian,opper2001adaptive} and its variants
\cite{fletcher2016expectation}.

Unfortunately, our results do not guarantee the convergence of the method 
to the fixed point.
However, in numerical experiments on sparse regression problems, we show that
the proposed method exhibits extremely stable convergence over a large
class of matrices that cause AMP to diverge.
Moreover, the performance of EM-VAMP is almost identical
to that of VAMP with known parameters.  In particular, the method is able
to obtain close to the theoretically optimal performance predicted by
the replica method~\cite{RanganFG:12-IT}.

\section{EM-VAMP}

\subsection{Review of VAMP}
To describe the VAMP method in \cite{rangan2016vamp},
we need to introduce some additional notation.  First suppose that we can write the prior
on $\xbf$ as
\beq \label{eq:pxf}
    p(\xbf|\thetabf_1) = \frac{1}{Z_1(\thetabf_1)}\exp\left[ -f_1(\xbf|\thetabf_1) \right],
\eeq
where $f_1(\cdot)$ is some penalty function and $Z_1(\thetabf_1)$ is a normalization constant.
We assume that $f_1(\cdot)$ is separable, meaning that
\beq \label{eq:f1sep}
    f_1(\xbf|\thetabf_1) = \sum_{n=1}^N f_{1n}(x_n|\thetabf_1),
\eeq
for scalar functions $f_{1n}$.  This corresponds to the case that, conditional
on $\thetabf_1$, $\xbf$ has independent components.
Also, we write the likelihood for the Gaussian model \eqref{eq:yAx} as
\begin{align} \label{eq:pyxf}
    p(\ybf|\xbf,\theta_2)
    &:= \frac{1}{Z_2(\theta_2)}
    \exp\left[-f_2(\xbf,\ybf|\theta_2)\right] \\
    f_2(\xbf,\ybf|\theta_2)
    &:= \frac{\theta_2}{2}\|\ybf-\Abf\xbf\|^2,
    \quad
    Z_2(\theta_2) = \left(\frac{2\pi}{\theta_2}\right)^{M/2}.
    \label{eq:f2guass}
\end{align}
The joint density of $\xbf,\ybf$ given parameters
$\thetabf=(\thetabf_1,\theta_2)$ is then
\beq \label{eq:pxy}
    p(\xbf,\ybf|\thetabf) = p(\xbf|\thetabf_1)p(\ybf|\xbf,\theta_2).
\eeq

The VAMP algorithm \cite{rangan2016vamp} considers the case where
the parameters $\thetabf$ are known.  In this case, VAMP attempts
to compute \emph{belief estimates} of the posterior density $p(\xbf|\ybf,\thetabf)$
of the form (for $i=1,2$)
\beq \label{eq:bidef}
    b_i(\xbf|\rbf_i,\gamma_i,\thetabf_i) \propto \exp\left[ -f_i(\xbf,\ybf|\thetabf_i) -
        \frac{\gamma_i}{2}\|\xbf-\rbf_i\|^2 \right],
\eeq
where the parameters $\rbf_i,\gamma_i$ are optimized by the
algorithm.  To keep the notation symmetric, we have written
$f_1(\xbf,\ybf|\thetabf_1)$ for $f_1(\xbf|\thetabf_1)$ even though the first penalty function
does not depend on $\ybf$.


The steps of VAMP are identical to those shown for
proposed EM-VAMP in Algorithm~\ref{algo:em-vamp},
except that VAMP skips the parameter updates
in lines~\ref{line:theta1} and \ref{line:theta2}.
Instead, VAMP fixes $\thetabfhat_{ik}$ for all iterations $k$.
In Algorithm~\ref{algo:em-vamp},
we have focused on the MMSE version of VAMP since we are interested in approximate
inference.  There we use
\[
    \Exp\left[ \phibf(\xbf) | \rbf_i,\gamma_i,\thetabf_i \right]
    := \int \phibf(\xbf)b_i(\xbf|\rbf_i,\gamma_i,\thetabf_i)\dif\xbf
\]
to denote the expectation with respect to the belief estimate $b_i(\cdot)$ in
\eqref{eq:bidef}.  Similarly, $\Cov(\cdot|\cdot)$ is the covariance matrix
with respect to the belief estimate and $\Tr \Cov(\cdot|\cdot)$ is its trace.
Hence, the VAMP method reduces the inference problem on the joint density
\eqref{eq:pxy} to computing expectations and variances with respect to
the belief estimates \eqref{eq:bidef}.

\begin{algorithm}[t]
\caption{EM-VAMP}
\begin{algorithmic}[1]  \label{algo:em-vamp}
\REQUIRE{Matrix $\Abf\in\R^{M\times N}$, penalty functions $f_i(\xbf,\ybf|\thetabf_i)$,
measurement vector $\ybf$,
and number of iterations $N_{\rm it}$.  }
\STATE{ Select initial $\rbf_{10}$, $\gamma_{10}\geq 0$, $\thetabfhat_{1,-1}$, $\thetahat_{2,-1}$.}
\FOR{$k=0,1,\dots,N_{\rm it}-1$}
    \STATE{// Input Denoising }
    \STATE{ $\thetabfhat_{1k} = \argmax_{\thetabf_1} \Exp\big[\ln p(\xbf|\thetabf_1)\big|\rbf_{1k},\gamma_{1k},\thetabfhat_{1,\km1} \big]$ }
        \label{line:theta1}
    \STATE{$\eta_{1k}^{-1} = (1/N)\Tr\big[ \Cov\big(\xbf\big|\rbf_{1k},\gamma_{1k},\thetabfhat_{1k}\big) \big]$}
        \label{line:eta1}
    \STATE{$\xbfhat_{1k} = \Exp\big(\xbf\big|\rbf_{1k},\gamma_{1k},\thetabfhat_{1k}\big)$}
        \label{line:x1}
    \STATE{$\gamma_{2k} = \eta_{1k} - \gamma_{1k}$}
        \label{line:gam2}
    \STATE{$\rbf_{2k} = (\eta_{1k}\xbfhat_{1k} - \gamma_{1k}\rbf_{1k})/\gamma_{2k}$}
        \label{line:r2}
    \STATE{ }
    \STATE{// LMMSE estimation }
    \STATE{ $\thetahat_{2k} = \argmax_{\theta_2} \Exp\big[\ln p(\ybf|\xbf,\theta_2)\big|\rbf_{2k},\gamma_{2k},\thetahat_{2,\km1} \big]$ }
        \label{line:theta2}
    \STATE{$\eta_{2k}^{-1} = (1/N)\Tr\big[ \Cov\big(\xbf\big|\rbf_{2k},\gamma_{2k},\thetahat_{2k}\big) \big]$}
        \label{line:eta2}
    \STATE{$\xbfhat_{2k} = \Exp\big(\xbf\big|\rbf_{2k},\gamma_{2k},\thetahat_{2k}\big)$}
        \label{line:x2}
    \STATE{$\gamma_{1,\kp1} = \eta_{2k} - \gamma_{2k}$}
        \label{line:gam1}
    \STATE{$\rbf_{1,\kp1} = (\eta_{2k}\xbfhat_{2k} - \gamma_{2k}\rbf_{2k})/\gamma_{1,\kp1}$}
        \label{line:r1}
\ENDFOR
\end{algorithmic}
\end{algorithm}

One of the main motivations of the VAMP method is that, for
the penalty functions \eqref{eq:f1sep} and \eqref{eq:f2guass} considered here,
the expectation and variance computations may be tractable at high dimensions.
To understand why, first observe that,
under the assumption of a separable penalty function \eqref{eq:f1sep},
the belief estimate $b_1(\cdot)$ separates as
\[
    b_1(\xbf|\rbf_1,\gamma_1,\thetabf_1) \propto \prod_{n=1}^N \exp\left[ -f_1(x_n|\thetabf_1) -
        \frac{\gamma_1}{2}(x_n-r_{1n})^2 \right].
\]
Thus, the expectation and variance computations in lines \ref{line:eta1} and \ref{line:x1}
decouple into $N$ scalar computations.
Furthermore, for the quadratic penalty \eqref{eq:f2guass}, the belief estimate $b_2(\cdot)$ is Gaussian, i.e.,
\[
    b_2(\xbf|\rbf_2,\gamma_2,\theta_2) \propto
    \exp\left[ -\frac{\theta_2}{2}\|\ybf-\Abf\xbf\|^2 - \frac{\gamma_2}{2}\|\rbf_2-\xbf\|^2
    \right] ,
\]
with mean and covariance given by
\begin{align}
    \Exp\left[\xbf|\rbf_2,\gamma_2,\theta_2\right]
    &=\Qbf^{-1}\left( \theta_2\Abf\tran\ybf + \gamma_2\rbf_2 \right) =: \xbfhat_2
    \label{eq:Exr2} \\
    \Cov\left[ \xbf | \rbf_2,\gamma_2,\theta_2 \right]
    &= \Qbf^{-1}
    \label{eq:Covxr2} \\
    \Qbf
    &= \theta_2 \Abf\tran\Abf + \gamma_2\Ibf .
\end{align}
Although \eqref{eq:Exr2}-\eqref{eq:Covxr2} may suggest that VAMP requires an $N\times N$ matrix inverse at each iteration, it is shown in \cite{rangan2016vamp} that two $M\times N$ matrix-vector multiplications per iteration are sufficient if the SVD of $\Abf$ is precomputed before initialization.
Thus, VAMP reduces the intractable posterior inference problem
to an iteration of $N$ scalar estimation problems and $2$ matrix-vector multiplies per iteration, just like AMP\@.

\subsection{Learning the parameters $\thetabf$}
To learn the parameters $\thetabf$, the EM-VAMP methods adds two steps,
lines \ref{line:theta1} and \ref{line:theta2}, to update $\thetabfhat_{ik}$.
These maximizations are similar to those in the EM method, and we formalize
this connection in the next section.  
The updates may be performed once per VAMP iteration, as written,
or several times per VAMP iteration, since in practice this seems to 
speed convergence of EM-VAMP\@.
For now, observe that due to the structure
of the prior in \eqref{eq:pxf} and the likelihood in \eqref{eq:pyxf},
we have that
\beq \label{eq:thetamin}
    \thetabfhat_{i,\kp1} = \argmin_{\thetabf_i}  \left\{
        \Exp\left[ f_i(\xbf,\ybf|\thetabf_i) \left|
        \rbf_{ik},\gamma_{ik},\thetabfhat_{ik} \right.\right]
        + \ln Z_i(\thetabf_i) \right\}.
\eeq
This minimization is often tractable.
For example, when the penalty function corresponds to an exponential family
(i.e., $f_i(\xbf,\ybf|\thetabf_i) = \thetabf_i\tran \phibf_i(\xbf,\ybf)$
for sufficient statistic $\phibf_i(\xbf,\ybf)$), the minimization
in \eqref{eq:thetamin} is convex.
In particular, for the quadratic loss \eqref{eq:f2guass},
the minimization is given by
\begin{align}
    \thetahat_{2,\kp1}^{-1} 
    &= \frac{1}{M}\Exp\big[
    \|\ybf-\Abf\xbf\|^2 \big| \rbf_{2k},\gamma_{2k},\theta_{2k} \big] \nonumber \\
    &=    \frac{1}{M} \left[
        \|\ybf-\Abf\xbfhat_{2k}\|^2 + \Tr(\Abf\Qbf_k^{-1}\Abf\tran ) \right]
    \label{eq:thetaminLMMSE},
\end{align}
where $\Qbf_k = \thetahat_{2k}\Abf\tran\Abf + \gamma_{2k}\Ibf$.
As mentioned earlier, it is possible to reduce the complexity of evaluating \eqref{eq:thetaminLMMSE} by precomputing the SVD of $\Abf$ \cite{rangan2016vamp}, 
since
$\Tr(\Abf\Qbf_k^{-1}\Abf\tran)
=\sum_{i=1}^R s_i^2/(\theta_{2k}s_i^2 + \gamma_{2k})$
where $\{s_i\}_{i=1}^R$ are the non-zero singular values of $\Abf$.
In this case, the update of $\theta_{2}$ is very simple, computationally.

\section{Fixed Points of EM-VAMP}


We will now show that
the parameter updates in EM-VAMP can be understood as
an approximation of the EM algorithm.  We first
briefly review the standard energy-function interpretation of EM 
\cite{NealHinton:98}.
Consider the problem of finding
the maximum likelihood (ML) estimate of the parameter $\thetabf$:
\beq \label{eq:thetaML}
    \thetabfhat = \argmax_{\thetabf} p(\ybf|\thetabf) = \argmax_{\thetabf} \int  p(\xbf,\ybf|\thetabf) \dif\xbf.
\eeq
Due to the integration, this minimization is generally intractable.
EM thus considers an auxiliary function,
\beq \label{eq:Qaux}
    Q(\thetabf,b) = -\ln p(\ybf|\thetabf) + D(b\|p(\cdot|\ybf,\thetabf)),
\eeq
defined for an arbitrary density $b(\xbf)$.
In \eqref{eq:Qaux},
$D(b\|p(\cdot|\ybf,\thetabf))$ is the KL divergence
between $b(\xbf)$ and the posterior density $p(\xbf|\ybf,\thetabf)$.
Note that, for any parameter estimate $\thetabf$,
\[
    \min_b Q(\thetabf,b) = -\ln p(\ybf|\thetabf),
\]
where the minimum occurs
at the posterior $\bhat(\xbf) = p(\xbf|\ybf,\thetabf)$.
Hence, the MLE \eqref{eq:thetaML} can, in principle,
be found from the joint minimization
\beq \label{eq:thetaminQ}
    \thetabfhat = \argmin_{\thetabf} \min_b Q(\thetabf,b).
\eeq
This fact leads to a natural alternating minimization,
\begin{align}
    \mbox{E-step:  } & \bhat_k = \argmin_b Q(\thetabfhat_k,b) = p(\xbf|\ybf,\thetabfhat_k) \label{eq:estep}
        \\
    \mbox{M-step:  } &\thetabfhat_{\kp1} = \argmin_{\thetabf} Q(\thetabf,\bhat_k). \label{eq:mstep}
\end{align}
This recursion is precisely the EM algorithm, written in a slightly non-standard form.  
Specifically,
\eqref{eq:estep} is the E-step, which computes the posterior density of $\xbf$
given $\ybf$ and the current parameter estimate $\thetabfhat_k$.  
A simple manipulation shows that
\beq \label{eq:Qaux1}
    Q(\thetabf,b) = -\Exp\left[ \ln p(\xbf,\ybf|\thetabf) | b \right] - H(b),
\eeq
where the expectation is with respect to the density $b(\xbf)$ and $H(b)$
is the differential entropy of $b$.  Equation~\eqref{eq:Qaux1} shows that the minimization in
\eqref{eq:mstep} can equivalently be written as 
\beq \label{eq:mstep1}
    \thetabfhat_{\kp1} = \argmax_{\thetabf}
    \Exp\left[ \ln p(\xbf,\ybf|\thetabf) \left| \bhat_k \right.\right],
\eeq
which is a familiar expression for the M-step.
Unfortunately, the computation of the posterior density required by the E-step
\eqref{eq:estep} is generally intractable for joint density \eqref{eq:pxy}
considered here.

We thus consider an alternate energy function, similar to that used
by Heskes in \cite{heskes2004approximate} for understanding EM combined with
belief propagation-based inference.
First observe that, using \eqref{eq:Qaux1}
and \eqref{eq:pxy}, we can write the auxiliary function as
\begin{align}
    Q(\thetabf,b) &= \sum_{i=1}^2 \left\{
        \Exp\left[ f_i(\xbf,\ybf|\thetabf_i) | b \right] + \ln Z_i(\thetabf_i) \right\}
        - H(b) \nonumber \\
        &= \sum_{i=1}^2 D_i(b,\thetabf_i) + H(b), \label{eq:Qaux2}
\end{align}
where $D_i(b,\thetabf_i)$ is the KL divergence,
\begin{align}
    D_i(b,\thetabf_i) = D\left(b\left\|Z_i(\thetabf_i)^{-1}e^{-f_i(\cdot,\ybf|\thetabf_i)}\right.\right)
    \label{eq:KLdiv}.
\end{align}
Now, given densities $b_1,b_2$ and $q$, we define the energy function 
\beq \label{eq:JBFE}
    J(b_1,b_2,q,\thetabf) := D_1(b_1,\thetabf_1) + D_2(b_2,\thetabf_2) + H(q),
\eeq
which matches the original auxiliary function $Q(\thetabf,b)$ under the matching
condition $b=b_1=b_2=q$.
Hence, we can rewrite the joint minimization \eqref{eq:thetaminQ} as
\beq\label{eq:JBFEOpt}
    \thetabfhat = \argmin_{\thetabf} \min_{b_1,b_2} \max_q J(b_1,b_2,q,\thetabf) \mbox{ s.t. }
    b_1=b_2=q.
\eeq
We call \eqref{eq:JBFE} the \emph{EM-VAMP energy function}.

Now, as mentioned in the Introduction,
VAMP---like many algorithms---can be viewed as an example of expectation consistent (EC) approximate inference \cite{opper2004expectation,OppWin:05,fletcher2016expectation}.
Specifically, following the EC framework, we relax the above GFE optimization
by replacing the constraints in \eqref{eq:JBFEOpt} with so-called
\emph{moment matching} constraints:
\begin{align} \label{eq:MMcon}
\begin{split}
    & \Exp(x_n|b_1) = \Exp(x_n|b_2) = \Exp(x_n|q), ~\forall n, \\
    & \Exp(\|\xbf\|^2 |b_1) = \Exp(\|\xbf\|^2 |b_2) = \Exp(\|\xbf\|^2 |q).
\end{split}
\end{align}
Thus, instead of requiring a perfect match in the densities $b_1,b_2,q$
as in \eqref{eq:JBFEOpt},
we require only a match in their first moments and average second moments.
Using the above approximation, we can then attempt to compute parameter estimates
via the minimization
\beq \label{eq:JBFEOptEC}
    \thetabfhat = \argmin_{\thetabf} \min_{b_1,b_2} \max_q J(b_1,b_2,q,\thetabf)
        \mbox{ s.t. \eqref{eq:MMcon} are satisfied}.
\eeq

Our main result shows that the fixed points of EM-VAMP are stationary
points of the optimization \eqref{eq:JBFEOptEC}.
To state the result, we write the Lagrangian
of the constrained optimization \eqref{eq:JBFEOptEC} as
\begin{align}
    L(b_1,b_2,q,\thetabf,\betabf,\gamma)
    &:= J(b_1,b_2,q,\thetabf)
        \!-\! \sum_{i=1}^2 \betabf_i\tran\left[
                \Exp(\xbf|b_i) \!-\! \Exp(\xbf|q) \right]
     \nonumber\\&\quad
     + \sum_{i=1}^2 \frac{\gamma_i}{2}\left[
        \Exp(\|\xbf\|^2|b_i) \!-\! \Exp(\|\xbf\|^2|q) \right],
    \label{eq:LagBFE}
\end{align}
where $\betabf=(\betabf_1,\betabf_2)$
and $\gamma=(\gamma_1,\gamma_2)$  represent sets of dual parameters
for the first- and second-order constraints.  We then have the following.

\begin{theorem} \label{thm:fix}
At any fixed point of the EM-VAMP algorithm with
$\gamma_1+\gamma_2 > 0$, we have
\begin{subequations}
\begin{align}
    \eta_1 &= \eta_2 = \eta := \gamma_1 + \gamma_2,
        \label{eq:etafix} \\
    \xbfhat_1 &= \xbfhat_2 = \xbfhat :=
        \left( \gamma_1\rbf_1 + \gamma_2\rbf_2 \right) / (\gamma_1+\gamma_2)
        \label{eq:xhatfix} .
\end{align}
\end{subequations}
Also, let $\betabf_i := \gamma_i\rbf_i$, let $\bhat_i$ be the density
\beq \label{eq:bhati}
    \bhat_i(\xbf) := b_i(\xbf|\rbf_i,\gamma_i,\thetabfhat_i),
\eeq
where $b_i(\cdot)$ is given in \eqref{eq:bidef}
and let $\qhat(\xbf)$ be the Gaussian density
\beq \label{eq:qfix}
    \qhat(\xbf) \propto \exp\left[ -\frac{\eta}{2}\|\xbf-\xbfhat\|^2 \right].
\eeq
Then, $\bhat_i$, $\thetabfhat$, and $\qhat$ are critical points of the Lagrangian 
\eqref{eq:LagBFE} that satisfy the moment matching constraints \eqref{eq:MMcon}.
\end{theorem}

\iftoggle{conference}{Due to space considerations, the proof is
given in the full paper \cite{fletcher2016emvamp}.
To summarize, the proof is an adaptation of
a similar result in \cite{fletcher2016expectation} with the addition of the
parameters $\thetabf$.}{
The proof is given in Appendix~\ref{sec:fixPf}
and is an adaptation of
a similar result in \cite{fletcher2016expectation} with the addition of the
parameters $\thetabf$.}
The consequence of this result is that, if the algorithm converges,
then its limit points are local minima of the EM-VAMP energy
minimization.

\begin{figure}[t]
\centering
\psfrag{(a)}{}
\includegraphics[width=0.95\columnwidth]{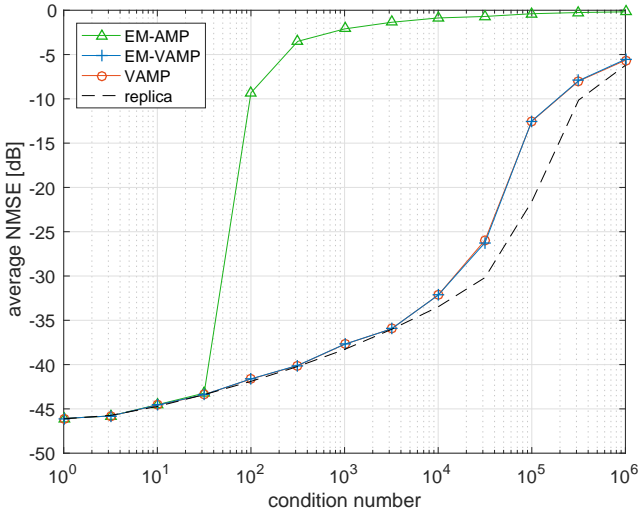}
\caption{For sparse linear regression, recovery NMSE versus condition number of $\Abf$.  Also shown is the replica prediction of the MMSE.}
\label{fig:vs_cond}
\end{figure}

\section{Numerical Experiments} \label{sec:sims}

While the above analysis characterizes the fixed points of EM-VAMP,
it does not provide any guarantees on the convergence of the algorithm to
the fixed points.
To study the convergence and evaluate the algorithm's performance,
we conducted a numerical experiment.

We considered sparse linear regression,
where the goal is to recover the signal $\xbf$ from measurements $\ybf$ from \eqref{eq:yAx} without knowing the signal parameters $\thetabf_1$ or the noise precision $\theta_2>0$.
For our experiment, we drew $\xbf$ from an
i.i.d.\ Bernoulli-Gaussian (i.e., spike and slab) prior,
\begin{equation}
p(x_n|\thetabf_1)=(1-\beta_x)\delta(x_n)+\beta_x{\mathcal N}(x_n;\mu_x,\tau_x)
\label{eq:BG},
\end{equation}
where parameters $\thetabf_1 = \{\beta_x,\mu_x,\tau_x\}$ represent the
sparsity rate $\beta_x\in(0,1]$, the active mean $\mu_x\in\R$, and the active variance $\tau_x>0$.
%
Following \cite{Vila:ICASSP:15},
we constructed $\Abf \in \R^{M \times N}$ from the singular value decomposition (SVD) $\Abf=\Ubf\Sbf\Vbf\tran$, whose orthogonal matrices $\Ubf$ and $\Vbf$ were drawn uniformly with respect to the Haar measure and whose singular values $s_i$ were constructed as a geometric series, i.e., $s_i/s_{i-1}=\alpha~\forall i>1$, with $\alpha$ and $s_1$ chosen to achieve a desired condition number
$s_1/s_{\min(M,N)}$ as well as $\|\Abf\|_F^2=N$.
It is shown in \cite{RanSchFle:14-ISIT,Vila:ICASSP:15} that standard AMP
(and even damped AMP) diverges when the matrix $\Abf$ has a sufficiently high
condition number.
Thus, this matrix-generation model provides an excellent test for the stability of AMP methods.
Recovery performance was assessed using normalized mean-squared error (NMSE) $\|\xbfhat-\xbf\|^2/\|\xbf\|^2$ averaged over $100$ independent draws of $\Abf$, $\xbf$, and $\wbf$.


\begin{figure}
\psfrag{(b)}[][][0.8]{\sf (a)}
\psfrag{(c)}[][][0.8]{\sf (b)}
\centering
\includegraphics[width=0.95\columnwidth]{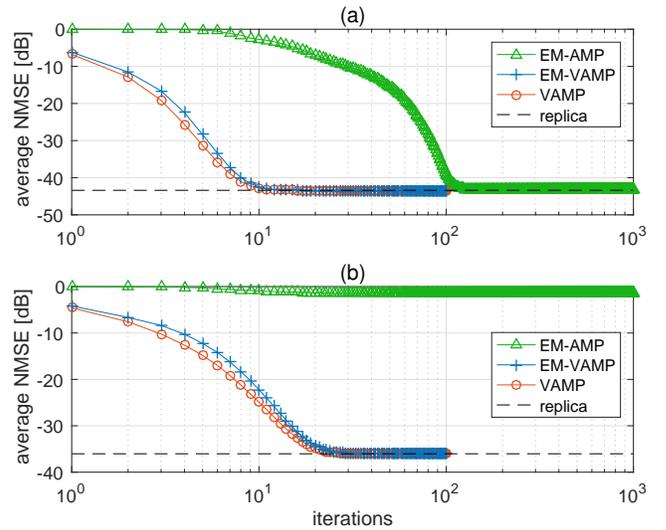}
\caption{For sparse linear regression, recovery NMSE versus iteration for condition number $32$ in (a) and condition number $3162$ in (b).}
\label{fig:vs_iter}
\end{figure}

Figure~\ref{fig:vs_cond} shows NMSE versus condition number for sparse linear regression under $M=512$, $N=1024$, $\beta_x=0.1$, $\mu_x=0$, and $(\tau_x,\theta_2)$ giving a signal-to-noise ratio of $40$~dB\@.
EM-VAMP was initialized with $\beta_x=(M/2)/N$, $\tau_x=\|\ybf\|^2 / \|\Abf\|_F^{2}\beta_x$, $\mu_x=0$, and $\theta_2^{-1}=M^{-1}\|\ybf\|^2$.
It is compared with (i) VAMP under perfect knowledge of $\thetabf=\{\tau_w,\beta_x,\mu_x,\tau_x\}$;
(ii) the EM-AMP algorithm from \cite{vila2013expectation} with damping from \cite{Vila:ICASSP:15}; and
(iii) the replica prediction for Bayes minimum MSE from \cite{tulino2013support}.
It was recently shown \cite{barbier2016mutual,reeves2016replica} that the replica method gives the correct prediction in sparse linear regression when $\Abf$ is i.i.d.\ Gaussian. 
Figure~\ref{fig:vs_cond} shows that the NMSE of EM-VAMP
is nearly indistinguishable from that of VAMP
and much more robust than EM-AMP to ill-conditioning in $\Abf$.

Figure~\ref{fig:vs_iter}(a) shows EM-VAMP and VAMP converging in $\sim 10$ iterations (whereas EM-AMP requires $>100$ iterations) at condition number $32$, and Figure~\ref{fig:vs_iter}(b) shows EM-VAMP converging in $\sim 20$ iterations at condition number $3162$.
These plots suggest that the convergence rate of EM-VAMP is
i) nearly identical to that of genie-aided VAMP and 
ii) relatively insensitive to the condition number of $\Abf$.
We note that, in generating the above figures, we used multiple updates of the noise precision $\theta_2$ per VAMP iteration.
In particular, \eqref{eq:thetaminLMMSE} was iterated to convergence.

A Matlab implementation of our EM-VAMP method can be found in the GAMPmatlab software package at\linebreak \url{http://sourceforge.net/projects/gampmatlab/}.

\section{Conclusions and Future Work}

We presented an approach for recovering the signal $\xbf$
from AWGN-corrupted linear measurements $\ybf = \Abf\xbf + \wbf$
by posing recovery in the MMSE framework while simultaneously
learning the parameters $\thetabf$ governing the signal prior
$p(\xbf|\thetabf)$ and the AWGN variance.
The proposed method combines EM and VAMP algorithms
for approximate inference of the posterior.
We showed that, if the algorithm converges, then
its fixed points coincide with stationary points of a certain
energy function.
Simulations show the proposed method exhibits robustness to the condition number of $\Abf$ and MMSE closely matching that of the replica prediction
under known $\thetabf$.

While the algorithm has great potential, one outstanding issue is that its
convergence has not been established.
One possible solution is to extend the convergence proofs in
\cite{fletcher2016expectation} or
the state evolution analysis of VAMP \cite{rangan2016vamp}.
%
%
Another avenue for future work is the application of EM-VAMP to \emph{sparse Bayesian learning} (SBL) \cite{Tipping:01}. 
SBL tackles sparse linear regression using a Gaussian-scale-mixture prior
$p(\xbf|\thetabf_1) = \mathcal{N}(\xbf;\zero,\Diag(\thetabf_1))$
with a deterministic unknown variance vector $\thetabf_1\in\R_+^N$ 
learned by the EM algorithm.
While the standard SBL implementation uses an $N\times N$ matrix inverse at each EM iteration, the EM-VAMP implementation of SBL could avoid matrix inversions by precomputing an SVD\@.

\iftoggle{conference}{\clearpage}{}
\bibliographystyle{IEEEtran}
\bibliography{../bibl}

\iftoggle{conference}{}{
\appendix

\subsection{Proof of Theorem~\ref{thm:fix}} \label{sec:fixPf}

The proof is modification of \cite{fletcher2016expectation}
with the addition of the parameters $\thetabf$.
From line~\ref{line:eta1} and \ref{line:eta2} of Algorithm~\ref{algo:em-vamp},
$\eta_i = \gamma_1 + \gamma_2$ for $i=1,2$, which proves \eqref{eq:etafix}.
Also, since $\gamma_1+\gamma_2>0$, we have that $\eta > 0$.
In addition, from lines~\ref{line:r2} and \ref{line:r1} we know,
\[
    \xbfhat_i = \left( \gamma_1 \rbf_1 + \gamma_2\rbf_2 \right)/\eta
    \text{~~for~~} i=1,2,
\]
which proves \eqref{eq:xhatfix}.

Now, by saying that $b_1,b_2,\qhat,\thetabfhat$ are fixed points of the Lagrangian, we mean that
\begin{align}
    \thetabfhat &= \argmin_{\thetabf} L(\bhat_1,\bhat_2, \qhat,\thetabf,\betabf,\gamma),
        \label{eq:thetafixlag} \\
    (\bhat_1,\bhat_2)
    &= \argmin_{b_1,b_2} L(b_1,b_2,\qhat,\thetabfhat,\betabf,\gamma), \label{eq:bifixlag} \\
    \qhat &= \argmax_{q} L(\bhat,q,\thetabfhat,\betabf,\gamma), \label{eq:qfixlag}
\end{align}
To prove \eqref{eq:thetafixlag}, first observe that, for $i=1,2$,
\begin{align}
     \lefteqn{ L(b_1,b_2,q,\thetabf,\betabf,\gamma) }\nonumber \\
     &= J(b,q,\thetabf) + \mbox{const} \nonumber \\
     &= \Exp\left[ f_i(\xbf,\ybf,\thetabf_i) | b_i \right]
        + \ln Z_i(\thetabf_i)+ \mbox{const},\label{eq:lagBFEt}
\end{align}
where the constant terms do not depend on $\thetabf$
and in the second step we used \eqref{eq:KLdiv} and \eqref{eq:JBFE}.
Using \eqref{eq:thetamin} and the definition of $\bhat_i(\xbf)$ in \eqref{eq:bifixlag},
we see that
\[
  \thetabfhat_{i,\kp1} = \argmin_{\thetabf_i}
        \Exp\left[ f_i(\xbf,\ybf|\thetabf_i) \left|\,
        \bhat_i \right.\right]   + \ln Z_i(\thetabf_i).
\]
Combining this with \eqref{eq:lagBFEt} establishes \eqref{eq:thetafixlag}.

To prove \eqref{eq:bifixlag}, we rewrite the
Lagrangian \eqref{eq:LagBFE} as
\begin{align}
    \lefteqn{ L(b,q,\theta,\betabf,\gamma) }\nonumber\\
        &\stackrel{(a)}{=}
        D(b_i \, \| \, e^{-f_i}) - \betabf_i\tran\Exp(\xbf|b_i) 
        + \frac{\gamma_i}{2}\Exp\left[ \|\xbf\|^2 |b_i \right] + \mbox{const}
         \nonumber \\
        &\stackrel{(b)}{=}
            D(b_i \, \| \, e^{-f_i})
            + \frac{1}{2}\Exp\left[\left. \gamma_i \|\xbf-\rbf_i\|^2 \right|b_i \right]
        + \mbox{const}
         \nonumber \\ &\stackrel{(c)}{=}
            -H(b_i) + \Exp\left[\left. f_i(\xbf,\ybf,\thetahat_i) + \frac{\gamma_i}{2}\|\xbf-\rbf_i\|^2
            \right|b_i \right]
        + \mbox{const} \nonumber \\
        &\stackrel{(d)}{=} D\left( b_i \left\| \, \bhat_i \right.\right) + \mbox{const},
        \label{eq:LagBFEbi2}
\end{align}
where in step (a) we removed the terms that do not depend on $b_i$;
in step (b) we used the fact that $\betabf_i = \gamma_i  \rbf_i$; and in steps
(c) and (d) we used the definitions of KL divergence and $\bhat_i$ in
\eqref{eq:bhati}.
Thus, the minimization in \eqref{eq:bifixlag} yields \eqref{eq:bhati}.

The maximization over $q$ in \eqref{eq:qfixlag} is computed
similarly.  Removing the terms that do not depend on $q$,
\begin{align}
    \lefteqn{ L(b_1,b_2,q,\betabf,\gamma) } \nonumber \\
        &= H(q) + \sum_{i=1}^2 \betabf_i\tran\Exp(\xbf|q)
        - \frac{\gamma_i}{2}\sum_{i=1}^2\Exp\left[ \|\xbf\|^2 |b_i \right] + \mbox{const}
        \nonumber \\
        &\stackrel{(a)}{=}
            H(q) + \eta \xbfhat\tran\Exp(\xbf|q)  
        - \frac{\eta}{2}\Exp\left[ \|\xbf\|^2 \right]   + \mbox{const}  \nonumber \\
        &\stackrel{(b)}{=}
            H(q) - \frac{\eta}{2}\Exp\left[ \|\xbf-\xbfhat\|^2 \right]   + \mbox{const}
            \nonumber \\
        & \stackrel{(c)}{=} -D(q \, \| \, \widehat{q} \,) + \mbox{const},
        \label{eq:LagBFEqi}
\end{align}
where step (a) uses the facts that $\gamma_1+\gamma_2=\eta$ and
\[
    \betabf_1 + \betabf_2 = \gamma_1 \rbf_1+\gamma_2 \rbf_2 = \eta \xbfhat,
\]
step (b) follows by completing the square, and step (c) uses the density
in \eqref{eq:qfix}.
Hence, the maximizer of \eqref{eq:qfixlag} is given by \eqref{eq:qfix}.

Also, from the updates of $\xbfhat_i$ and $\eta_i$ in Algorithm~\ref{algo:em-vamp},
we have
\[
    \xbfhat = \Exp(\xbf|b_i), \quad \eta^{-1} = \frac{1}{N}\Tr\Cov(\xbf|b_i).
\]
Since $\widehat{q}$ is Gaussian, its mean and average covariance are
\[
    \Exp(\xbf|q) = \xbfhat, \quad \frac{1}{N} \Tr\Cov(\xbf|q) = \eta^{-1}.
\]
This proves that the densities satisfy the moment matching constraints \eqref{eq:MMcon}.


}

\end{document}